\documentclass[jcp,twocolumn,showpacs,floats]{revtex4}
\usepackage{graphicx}
\usepackage{amsmath}
\usepackage{tikz}
\usepackage{color}
\graphicspath{{images/},{images/pdf/}}
\begin{document}
%-------------------------------------------------------------------------------
\title{Adsorption and ring-opening of lactide on a chiral metal surface studied by Density Functional Theory}
\author{J.-H.~Franke} 
\author{D.S. Kosov}
\altaffiliation{College of Science, Technology and Engineering, James Cook University,Townsville, QLD, 4811, Australia}
\affiliation{Department of Physics, Campus Plaine - CP 231, Universite Libre de Bruxelles, 1050 Brussels, Belgium}
\date{\today}

\begin{abstract}
We study the adsorption and ring-opening of lactide on the naturally chiral metal surface Pt(321)$^S$. Lactide is a precursor for polylactic acid ring-opening polymerization and Pt is a well known catalyst surface. We study here the energetics of the ring-opening of lactide on a surface that has a high density of kink atoms. These sites are expected to be present on a realistic Pt surface and show enhanced catalytic activity. The use of a naturally chiral surface also enables us to study potential chiral selectivity effects of the reaction at the same time. Using Density Functional Theory (DFT) with a functional that includes the van der Waals forces in a first-principles manner, we find modest adsorption energies of around 1.4 eV for the pristine molecule and different ring-opened states. The energy barrier to be overcome in the ring-opening reaction is found to be very small at 0.32 eV and 0.30 eV for LL- and its chiral partner DD-lactide, respectively. These energies are much smaller than the activation energy for a dehydrogenation reaction of 0.78 eV. Our results thus indicate that (a) ring-opening reactions of lactide on Pt(321) can be expected already at very low temperatures and Pt might be a very effective catalyst for this reaction; (b) the ring-opening reaction rate shows noticeable enantioselectivity.
\end{abstract}
\maketitle
%-------------------------------------------------------------------------------

\section{Introduction}

\begin{table*}[Htb]
\caption{Adsorption energies of the two enantiomers of lactide on Pt(321)$^S$. Numbers is brackets denote the contributions of the non-local correlation energy to the adsorption energy.}
\begin{tabular}{c|c|c}
\hline
\multicolumn{1}{c|}{molecule}& \multicolumn{1}{|c|}{configuration} & \multicolumn{1}{|c|}{adsorption energy oPBE-vdW (eV)}  \\
\hline
                       & facet-bridging 1 & -1.38 (-1.71)  \\
pristine LL-lactide    & facet-bridging 2 & -1.32 (-1.82)  \\
                       & ridge            & -1.25 (-1.72)  \\
\hline
                       & facet-bridging 1 & -1.37 (-1.85)  \\
pristine DD-lactide    & ridge            & -1.26 (-1.80)  \\
                       & facet-bridging 2 & -1.20 (-1.57)  \\
\hline
ring-opened LL-lactide & kink-ridge       & -1.41 (-1.99)  \\
                       & kink-kink        & -1.39 (-1.63)  \\
\hline
ring-opened DD-lactide & kink-ridge       & -1.42 (-1.95)  \\
                       & kink-kink        & -1.39 (-1.52)  \\
\hline
\hline
LL-lactide             & ring-opening transition state     & -1.06 (-2.11) \\
DD-lactide             & ring-opening transition state     & -1.07 (-2.18) \\
\hline
LL-lactide             & dehydrogenated                    & -1.33 (-2.00) \\
                       & dehydrogenation transition state  & -0.60 (-1.96) \\
\hline
\end{tabular}             
\label{tab:energ}
\end{table*}

Polylactic acid (PLA) has garnered much attention because it is biodegradable, bioabsorbable and can also be biologically-derived.\cite{Rasal2010,Platel2008} Its production consumes carbon dioxide and is also relatively energy efficient. PLA is compostable and is more processable than other biopolymers. Nevertheless, its most important strength has historically been its biocompatibility and biodegradability that is exploited by implants able to degrade inside the body into safe products.\cite{Gross2002,Rasal2010,Platel2008} Direct polymerization of the monomer, lactic acid, produces mostly modest molecular weight polymer and exhibits a long polymerization time.\cite{Platel2008,MadhavanNampoothiri2010} Therefore, the usual way to produce high-molecular weight PLA is by first condensing lactic acid, removing the condensation water, then breaking down the polymer into lactide. This is then purified and subsequently used as the starting point to form PLA via ring-opening polymerization.\cite{Platel2008,MadhavanNampoothiri2010,Katiyar2010,Dechy-Cabaret2004} Several catalysts are already known for solution and bulk polymerization, and research into new ones is ongoing.\cite{Platel2008,Katiyar2010,Dechy-Cabaret2004,Dijkstra2011,Thomas2010,Chen2012,Liu2012,Stopper2012,Saridis2013,Thomas2014,Bouyahyi2011,Marshall2005,Bonduelle2008,Dyer2010} To fundamentally understand the system, however, it is also worthwhile to look at it from the angle of surface science and to study how ring-opening polymerization works on metal surfaces.

Lactide and Pt(321) are chiral, which gives us an opportunity to study enantioselective catalytic chemical reactions. Chirality is ubiquitous in nature and has important consequences as the two mirror images, or enantiomers, can have very different interactions with other chiral molecules.\cite{Sholl2009,Gellman2010} It is thus desirable to obtain molecules of controlled chirality. To this end, one has to rely again on chirally specific interactions, to either separate one enantiomer from a racemic mixture or to preferrably catalyse the production of only one enantiomer by chirally specific catalysts.\cite{Rekoske2001,Blaser2005} Often, chirally selective reactions are carried out in solution, which necessitates the separation of product and catalyst after the reaction.\cite{Blaser2005,Sholl2009} This could be avoided by reacting the molecules on a surface that, in order to exhibit chiral selectivity, needs to be chiral itself. Fixing chiral molecules to a surface is a straightforward way to achieve this.\cite{Lorenzo2000,Fasel2006,Kuhnle2002,Lawton2013,Mallat2007,Meemken2012,Gross2013,Kyriakou2011} However, simple metal crystals can also provide chiral surfaces if they are cut in certain ways.\cite{Sholl2009,Mallat2007,Ahmadi1999,Sholl1998,Clegg2011,Kyriakou2011} This provides for simple, well-controlled experimental conditions that can be used to understand fundamentally how chiral surfaces interact with different enantiomers of a chiral molecule.\cite{Eralp2011,Bombis2010,Zhao2004,Greber2006,Horvath2004,Huang2011,Huang2008,Cheong2011,Han2012,Bhatia2005,Bhatia2008,Sljivancanin2002,Schmidt2012,James2008,Yun2014} In the case of PLA, its thermochemical properties depend on the chirality of its constituents, which therefore provides an additional degree of freedom to tune polymer properties.\cite{MadhavanNampoothiri2010,Dijkstra2011}

Here we focus on the ring-opening reaction of lactide on Pt surfaces. As under-coordinated surface atoms are often considered active catalytic sites, we adsorb the molecule on a surface with a high density of such kink atoms, Pt(321). Additionally, this surface is intrinsically chiral, thus allowing to check for eventual chiral selectivity effects on this reaction at the same time. Specifically, we identify the most stable adsorption configurations of LL-lactide and DD-lactide on Pt(321)$^S$, in its pristine and ring-opened forms and find that the adsorption energy is very similar in both states. We also calculate the transition states to quantify the energy barriers between the two states. We find that the transition state energy shows chiral selectivity with values of 0.32 eV (0.30 eV) for LL-(DD-)lactide. This is also much lower than the energy barrier calculated for a dehydrogenation at the methyl group (0.78 eV) that was calculated for comparison. Our results therefore indicate that ring-opening on kink sites of Pt surfaces is very likely to occur already at low temperatures. We find evidence of chiral selectivity as the energy barrier as well as adsorption energies of the product slightly favor the reaction of DD-lactide.

The remainder of the paper is organized as follows. First we outline our computational approach in Section 2. After that we discuss the adsorption of pristine lactide in Section 3 and of the ring-opened variety in Section 4. In Section 5 we discuss our findings for the transition states of the ring-opening reaction and the dehydrogenation of the methyl group.

\section{Computational details}

\begin{figure*}[Htb]
\includegraphics[width=13cm]{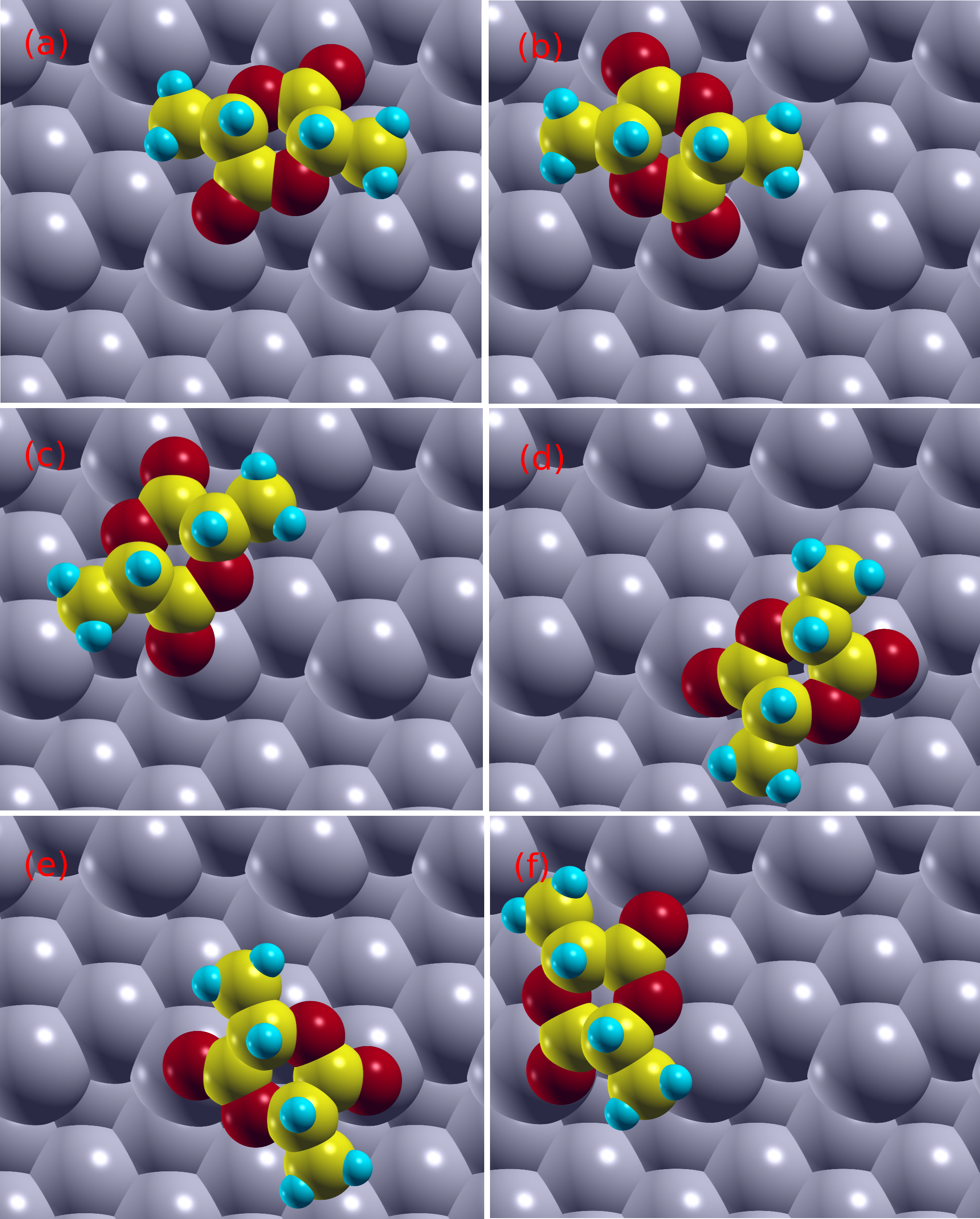}
\caption{LL-lactide (\textbf{a},\textbf{c},\textbf{e}) and DD-lactide (\textbf{b},\textbf{d},\textbf{f}) adsorbed on Pt(321)$^S$. The 3 most stable configurations are given in order of decreasing stability.}
\label{fig:im1}
\end{figure*}

We obtained our results with the DFT code VASP 5.3.2\cite{Hafner2008,Kresse1996a,Kresse1996b} with the oPBE-vdW functional\cite{Klimes2010,Klimes2011} throughout. The inclusion of the van der Waals forces is crucial for the adsorption of a weakly bound molecule since they dominate the binding energy.\cite{Franke2013,Franke2013a,Wei2014} We opted for this special version of the vdW-DF functional since we wanted to keep the PBE class of functionals while also aiming at an optimal accuracy with the vdW nonlocal correlation. The Projector Augmented Wave method\cite{Bloechl1994,Kresse1999} was employed with valence wave functions expanded up to an energy cutoff of 400 eV. All structural relaxations are carried out until all forces are smaller than 10 meV/\AA\ with wavefunctions converged to energy changes between successive steps smaller than 10$^{-5}$ eV. For all slab calculations dipole corrections to the potential are applied throughout.\cite{Neugebauer1992}

The (321) surfaces were constructed with a thickness corresponding to 28 layers of (321) orientation with the upper 14 layers of the slabs free to relax. Lactide was adsorbed on the relaxed side of a 3x2 supercell of the surface which constitutes a (321)$^S$ surface.\cite{Ahmadi1999,Sholl2001} A K-mesh of 3x3x1 was used in conjunction with Gaussian broadening of 0.1 eV. All molecular degrees of freedom were allowed to relax as were the upper 14 layers of the metal slabs. Adsorption energies $E_{adsorption}$ are given with reference to the energy of the isolated surface $E_{surface}$ before adsorbing the molecule, using identical computational parameters and supercell and the energy of the molecule in a large vacuum supercell $E_{lactide}$.

\begin{equation}
E_{adsorption} = E_{mol\ on\ surface} - E_{surface} - E_{lactide}.
\end{equation}

The transition state searches were carried out using initially the nudged elastic band (NEB) method starting from a linear interpolation between starting and end states.\cite{Sheppard2008,Henkelman2000,Henkelman2000a} For distances over 10 \AA\ between the starting and end states, a single image was relaxed first to trace out the molecular diffusion that makes up most of this distance. The continuity of the band was ensured by relaxing the image back to the ground state, thus tracing out part of the reaction coordinate. NEB calculations were then continued using 4-5 images between the obtained image and the stable state other from the one it relaxed too. In this way the distances between neighboring images could be kept at or below 2 \AA\ during the relaxation. Initially, several hundred ionic steps were carried out to equilibrate the distances between the images. Afterwards, the run was continued using the image with the highest energy as climbing image.\cite{Henkelman2000} In cases where the climbing image pulled heavily on one side of the chain, increasing the distance between neighboring images above 2 \AA, additional images were interpolated to ensure adequate accuracy of the reaction coordinate for the climbing image. During the relaxation, we repeatedly started dimer runs to see if they relax to an imaginary mode that corresponds to the desired reaction.\cite{Kastner2008,Heyden2005,Olsen2004,Henkelman1999} The final transition state was controlled by relaxing structures interpolated between it and the starting and final states of the reaction.

The NEB and Dimer calculations were carried out using the VTST code in version 3.0d. The spring constant of the NEB calculations was 5 eV/\AA$^2$ throughout and structural optimization was carried out with the FIRE optimizer\cite{Bitzek2006} with initial and maximum time steps of 0.08 and 0.5, respectively. The maximum displacement between successive steps was limited to 0.1 \AA. The wavefunctions were converged to energy changes below 10$^{-4}$ eV during the NEB calculations. For the Dimer calculations a much higher accuracy of the wave functions was used to accurately find the negative curvature mode (energy changes smaller than 10$^{-7}$ eV). The calculation was considered converged when forces were smaller than 10 meV/\AA\ on all ions.

\section{LL- and DD-Lactide on Pt(321)}

\begin{figure*}[Htb]
\includegraphics[width=13cm]{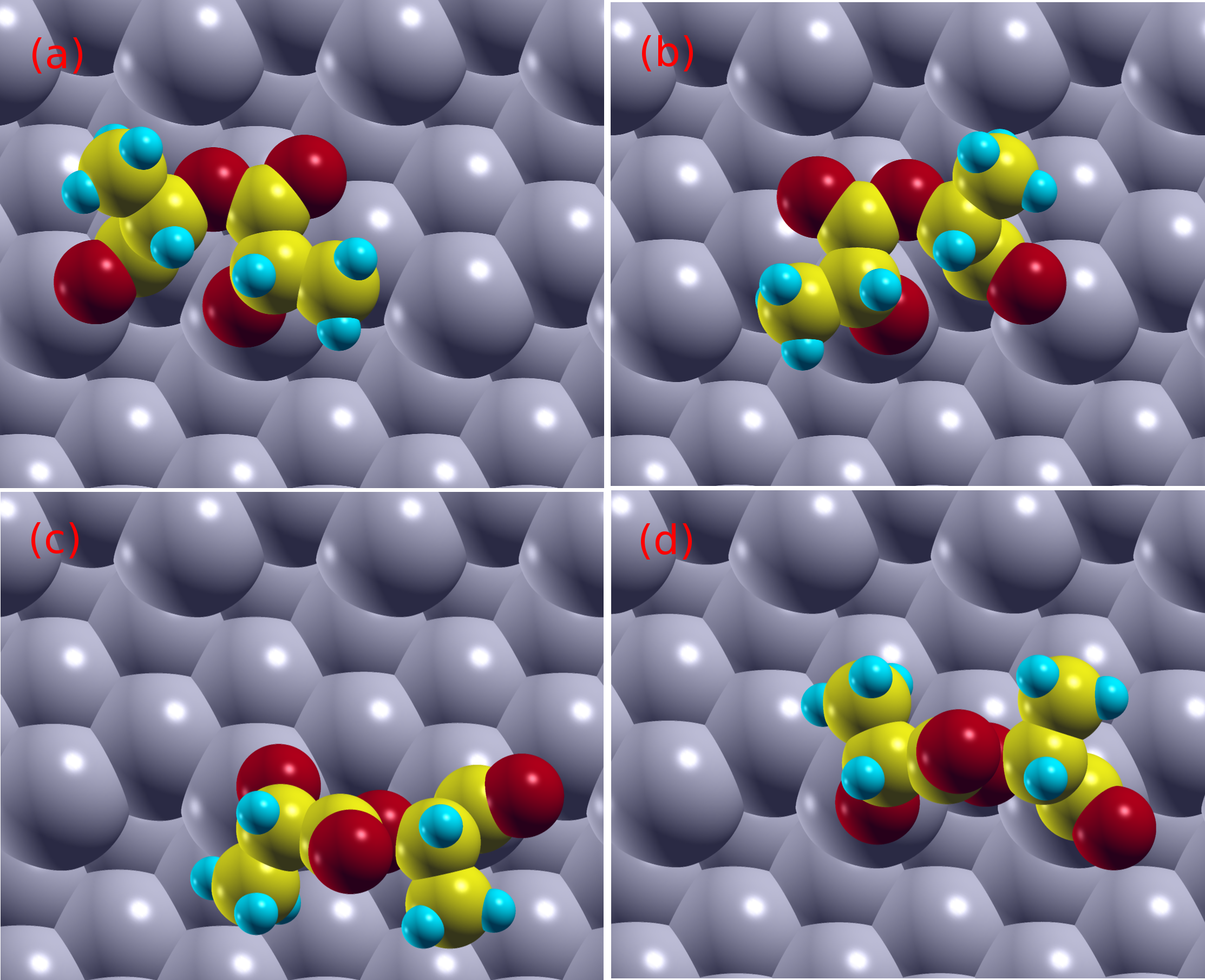}
\caption{Ring-opened LL-lactide (\textbf{a},\textbf{c}) and DD-lactide (\textbf{b},\textbf{d}) adsorbed on Pt(321)$^S$. The 2 most stable configurations are given in order of decreasing stability.}
\label{fig:im2}
\end{figure*}

\begin{table*}[Htb]
\caption{Bond lengths ( in \AA) for different conformations of lactide on Pt(321)$^S$. Pt $\cdots$ O=C refers to the interaction of the carbonyl oxygen with the nearest Pt atom of the surface. Pt-C and Pt-O refers to the bond lengths of the Pt atoms saturating the dangling bonds resulting from ring-opening. C-O is the bond length of the C-O bonds in the dioxane ring. Two numbers are given for the intact molecule and only the number of the opened bond for all transition state and ring-opened configurations. Finally, Pt-H and Pt-C give the bond lengths of the substrate bonds formed during dehydrogenation of the methyl group.}
\begin{tabular}{c|c|c|c|c|c|c|c}
\hline
\multicolumn{1}{c|}{molecule}& \multicolumn{1}{|c|}{configuration} & \multicolumn{1}{|c|}{Pt $\cdots$ O=C} & \multicolumn{1}{|c|}{Pt-C} & \multicolumn{1}{|c|}{Pt-O} & \multicolumn{1}{|c|}{C-O} & \multicolumn{1}{|c|}{Pt-H} & \multicolumn{1}{|c|}{Pt-C}\\
\hline
                       & facet-bridging 1                  & 2.19, 3.21 &      &      & 1.37, 1.33 &      &  \\
pristine LL-lactide    & facet-bridging 2                  & 2.24, 2.48 &      &      & 1.34, 1.35 &      &  \\
                       & ridge                             & 2.31, 2.34 &      &      & 1.35, 1.35 &      &  \\
\hline
                       & facet-bridging 1                  & 2.25, 2.39 &      &      & 1.35, 1.34 &      &  \\
pristine DD-lactide    & ridge                             & 2.22, 2.37 &      &      & 1.35, 1.34 &      &  \\
                       & facet-bridging 2                  & 3.62, 2.21 &      &      & 1.37, 1.33 &      &  \\
\hline
ring-opened LL-lactide & kink-ridge                        & 2.23       & 1.96 & 2.01 & 2.93       &      &  \\
                       & kink-kink                         & 3.32       & 1.99 & 1.98 & 4.83       &      &  \\
\hline
ring-opened DD-lactide & kink-ridge                        & 2.20       & 1.96 & 2.02 & 2.81       &      &  \\
                       & kink-kink                         & 3.97       & 1.99 & 1.98 & 4.88       &      &  \\
\hline
\hline
LL-lactide             & ring-opening transition state     & 2.15       & 2.06 & 2.17 & 1.78       &      &      \\
DD-lactide             & ring-opening transition state     & 2.14       & 2.07 & 2.20 & 1.68      &      &      \\
\hline
LL-lactide             & dehydrogenated                    & 2.22       &      &      &            & 1.71 & 2.09 \\
                       & dehydrogenation transition state  & 2.23       &      &      &            & 1.61 & 1.52 \\
\hline
\end{tabular}             
\label{tab:bonds}
\end{table*}

\begin{figure*}[Htb]
\includegraphics[width=13cm]{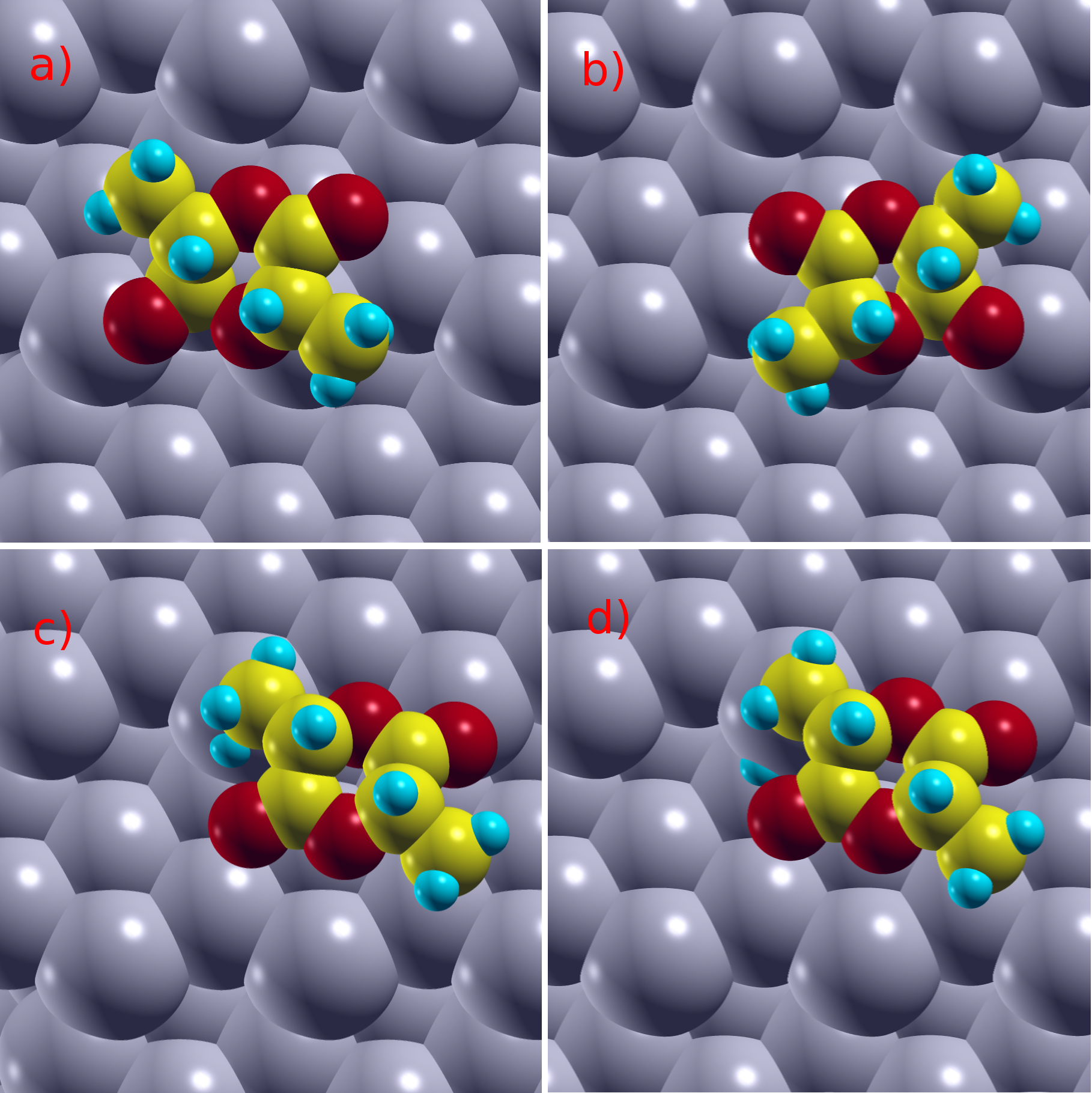}
\caption{Transition states of ring-opening of LL-lactide (\textbf{a}) and DD-lactide (\textbf{b}) on Pt(321)$^S$. Also shown are the transition state for the dehydrogenation of the methyl group (\textbf{c}) and the end state of this reaction (\textbf{d}).}
\label{fig:im3}
\end{figure*}

To find the most stable adsorption configurations of lactide on the Pt(321) surface, we built on previous results that oxygen atoms interact preferably with the kink atoms.\cite{Franke2013,Franke2013a} For lactide there is oxygen in the dioxane ring and as carbonyl substituents. We therefore generated test configurations with either of the two species above a kink site and rotation of the molecule in steps of 60 degrees. We did this for both LL-lactide and DD-lactide on Pt(321)$^S$ (see Table \ref{tab:bonds}.

For most adsorption configurations mutiple oxygen atoms came to interact with kink sites due to their high density on this specific surface. It turns out that the configurations that have carbonyl oxygen interacting with the kink sites are the most stable ones, while the dioxane ring oxygen interacts more weakly. In fact, all adsorption sites generated from the carbonyl oxygen atoms bound to kink sites relaxed to configurations with both carbonyl oxygen atoms bound to kink sites. All of these configurations are more stable than the most stable one generated with the ring atom bound to a kink site. The difference between the carbonyl-kink bound configurations amounts to differences in the two neighboring kink sites that the second carbonyl oxygens binds to. There are six possible binding combinations to nearest kink sites, of which only three are symmetry inequivalent due to the C2 rotational symmetry around the center of mass of the lactide molecule. Of these three, two are configurations where the lactide bridges the facet, and one where the molecule attaches with both carbonyl oxygen atoms to the same ridge (cf. Fig. \ref{fig:im1}).  

The most stable of the three configurations turns out to be a facet bridging configuration, with both methyl groups over the same facet, for both LL- and DD-lactide (see Table \ref{tab:energ}). For LL-lactide, the second most stable configuration is the other bridging configuration, about 0.06 eV higher in energy. The least stable of the three is the one with the molecule bound along the ridge, which is 0.13 eV higher in energy than the most stable one. For DD-lactide the ranking in energy of the second and third most stable configurations is opposite to the one calculated for LL-lactide, with adsorption energies 0.11 eV and 0.17 eV above the most stable configuration. The binding energies are similar to the binding energies obtained for a single lactic acid molecule on Pt(321).\cite{Franke2013} Also, the non-local correlation energy part of the binding energy is larger than the total binding energy, showing the importance of the van der Waals forces to the binding of lactide on Pt(321). 

The bond lengths of the carbonyl oxygens to the kink atoms of the surface vary between 2.19 \AA\ and 2.48 \AA\ for all configurations, except for one oxygen atom that is unbound (bond length of 3.64 \AA) in the facet-bridging 2 configuration of DD-lactide.

\section{Ring-opened lactide on Pt(321)}

To determine if the ring-opening reaction might occur on Pt(321) we found the most stable adsorption configurations of ring-opened versions of LL- and DD-lactide. We assume that the dangling bonds that result from the ring-opening bind to neighboring Pt atoms on a ridge as a result of the ring-opening reaction being catalyzed by a kink atom. Without putting too much strain on the molecule, the possible combinations are one dangling bond binding to a kink atom and the other one either to the neighboring ridge atom or the next kink atom. We refer to these configurations as kink-ridge and kink-kink, respectively. For the kink-ridge configurations there is the possibility to bind either the carbon or the oxygen dangling bond to the kink atom. For all three possibilities, two configurations related to each other by a rotation of the molecule by 180 degrees exist. So, overall, 6 configurations were calculated for each lactide enantiomer.

The two most stable configurations for each enantiomer are given in Table \ref{tab:energ} and Fig. \ref{fig:im2}. For both enantiomers, the kink-ridge conformation with the carbon atom bound to the ridge atom is the most stable, with the more open kink-kink conformation being less stable by an insignificant 0.02 - 0.03 eV. This small energy difference indicates that the molecule might be able to transfer back- and forth between the two states at an appropriate temperature. Interestingly, the energy of the ring-opened lactide on Pt(321) is also very close to the energy of the pristine molecule, being more stable by 0.03 eV and 0.05 eV for ring-opened LL- and DD-lactide, respectively. 

For all configurations the dangling bonds from the ring-opening bind to surface atoms with small bond lenghts, indicating strong bonds. The Pt-C bonds exhibit bond lenghts of 1.96 \AA\ and 1.99 \AA\ for the kink-ridge and kink-kink configurations, respectively. The Pt-O bond lengths are similar with 2.01 \AA\ (2.02 \AA\ for DD-lactide) and 1.98 \AA\ for the kink-ridge and kink-kink configurations, respectively. For the kink-ridge conformations, additionally one carbonyl oxygen interacts with a kink atom with bond distances of 2.23 \AA\ for LL-lactide and 2.20 \AA\ for DD-lactide. For the kink-kink configurations this bond is not formed, possibly contributing to its slightly lower stability. The distances between the carbon and oxygen atoms that were forming the ring are 2.93 \AA\ (2.81 \AA) for the kink-ridge conformation and 4.83 \AA\ (4.88 \AA) for the kink-kink conformation of LL-lactide (DD-lactide). Taken together, the kink atoms provide an ideal binding site for the ring-opened molecule that stabilizes the dangling bonds at a small distance. Additionally, in the case of the kink-kink binding configurations it is stabilized by a carbonyl oxygen-Pt bond.

\section{Ring-opening and dehydrogenation}

After establishing that initial and final states of a ring-opening reaction of lactide on Pt(321) are similar in energy, we calculated the transition state (TS) energy between the two states for both LL- and DD-lactide. To test if this reaction could indeed be observed on the surface, we also calculated a dehydrogenation at the methyl group as a competing process. We chose the dehydrogenation since it is often found in on-surface coupling of molecules and therefore seems a natural choice out of a multitude of possible bond-breaking events.\cite{Petersen2004,Petersen2004a,Anghel2005,Anghel2005a,Anghel2007,Zhong2011,Gao2013,Bebensee2013,Bjork2014,Vajda2009}

We found that the TS energy is very low for the ring-opening of LL-lactide at about 0.32 eV above the energy of the most stable adsorption configuration of the intact molecule and 0.35 eV above the energy of the ring-opened final state.  A temperature of 150K would give a Boltzmann factor of 10$^{-11}$ which would, in a transition state theory framework with a realistic trial rate, translate to an observable reaction rate. In comparison, the TS for the dehydrogenation of LL-lactide is calculated as 0.78 eV above the most stable adsorption configuration, which would give a Boltzmann factor of 10$^{-28}$ at 150K. Thus, our data indicate that the ring-opening reaction is much easier than the dehydrogenation and, under the assumption that the dehydrogenation is a good proxy for other possible reactions, might be observed at very low temperatures.

There is a small difference in the TS energy for the ring-opening of LL-lactide and DD-lactide of 0.02 eV.  Additionally, the energy difference between initial and final states of the reaction is 0.02 eV larger (-0.03 eV and -0.05 eV for LL-lactide and DD-lactide, respectively). This very subtle energy differences might lead to differences in the ratio of ring-opened to pristine molecules for the two enantiomers and indicate a chiral selectivity, because an 0.02 eV energy difference corresponds to ratios of the Boltzmann factors for the two enantiomers of about 1:5 at 150K.

Interestingly, the non-local correlation part of the binding energy (cf. Table \ref{tab:energ}) is larger for the ring-opening TSs than for all other configurations calculated. It stabilizes the molecule by 0.4 eV (0.33 eV) for LL- (DD)-lactide, when compared to the initial state of the reaction. The non-local correlation binding energy component is also 0.07 eV larger for the TS of DD-lactide than for the corresponding state of LL-lactide. This difference is thus larger then the calculated difference between the two enantiomers showing that an optimization of the non-local binding energy component for DD-lactide leads to our calculated chiral selectivity in the transition states.

At the TS, the bond length of the C-O bond that is eventually broken is extended from 1.37 \AA\ to 1.78 \AA\ in the case of LL-lactide (see Table \ref{tab:bonds}. At the same time, the three molecule-surface bonds that are eventually formed (Pt-C, Pt-O and the bond of the carbonyl oxygen) are already near their equilibrium value with bond lenghts of 2.06 \AA, 2.17 \AA\ and 2.15 \AA, respectively. For DD-lactide the molecule-surface bond lengths are very similar, but the C-O bond is broken already at 1.68 \AA. This smaller bond length at the TS is correlated with the smaller distance in the final state of the reaction. For the dehydrogenation, the C-H bond length is extended to 1.52 \AA\ and the bonds formed during the reaction, i.e. the Pt-C bond and the Pt-H bond are 2.24 \AA\ and 1.61 \AA\ long, respectively.

\section{Conclusion}

We identified the most stable adsorption geometries of LL- and DD-lactide on Pt(321)$^S$. The molecules are most stable in a configuration bridging the {111} oriented facets with the carbonyl oxygen binding to kink atoms. Ring-opened lactide was found to bind most strongly with its dangling bonds to a kink atom and a neighboring atom on the ridge. Configurations with both dangling bonds bound to kink atoms were slightly less stable. 

The calculated adsorption energy is about 1.4 eV without any obvious chiral selectivity. The ring-opened variety of the molecule is marginally more stable with energy differences between the pristine and the ring-opened molecule of only 0.03 eV and 0.05 eV for LL- and DD-lactide, respectively. A similarly small energy difference is obtained between the different binding patterns of the ring-opened lactide molecule. For all adsorption configurations, the contribution of van der Waals forces to the binding is dominant, as the non-local part of the correlation energy provides more than 100\% of the overall binding energy.

Energy barriers for the ring-opening reaction are very low at 0.32 eV (LL-lactide) and 0.30 eV (DD-lactide) which would make this reaction observable already at low temperatures, assuming realistic prefactors in a transition state theory framework. This small barrier can be understood from the structural similarity of the transition state and the ring-opened state of the molecule. As a result, the stabilizing molecule-surface interactions of the final state can form simultaneously to the breaking of the C-O bond. The specific geometry of the kink sites found on the Pt(321) surface with one ridge atom between two kink atoms seems to be ideal for the ring-opening reaction of lactide. Importantly, this configuration is not as peculiar as it might seem: it should be a feature of every rounded step edge on any Pt surface vicinal to the {111} direction. Highly active catalytic sites should therefore be present on real, roughened Pt surfaces.

The small energy changes between different ring-opened and pristine states of the molecule points to a sizable population of different configurations already at low temperature. The small transition state energy should hereby enable the seamless conversion of ring-opened to pristine molecular states. This structural variability should be important to actually connect two ring-opened molecules, a reaction step that is beyond the scope of this study due to the size of the system. Interestingly, the small difference in energy gain due to ring-opening between LL- and DD-lactide as well as the slightly smaller energy barrier might lead to a subtle difference between ring-opened LL- and DD-lactide molecules that might impact polymerization rates. Also, the second facet-bridging configuration of pristine LL-lactide is significantly more stable than the DD-lactide configuration. This might impact diffusion rates as well as the free energy of the molecule on the surface at finite temperatures. The first effect would decrease the ratio of LL- to DD- polymerization rates by increasing the likelihood of finding a non-opened lactide molecule at the end of the chain. The second effect would increase it by providing a better supply of molecules to the chain end. Also noteworthy is the different bond length at which the transition state for the ring-opening occurs. One of the two might be preferable at different average positions of the kink atoms on real, roughened Pt surfaces at finite temperatures.

The energy barrier for dehydrogenation is significantly larger than for the ring-opening at 0.78 eV. These data indicate that Pt surfaces vicinal to the {111} direction are a possible catalyst for ring-opening polymerization of lactic acid to poly-lactic acid. There are signs of enantiomeric selectivity of the reaction, with subtle differences in stable state energies and the energy barrier, together with differences in geometries.

\section{Acknowledgements}

This work has been supported by the Francqui Foundation, and Programme d'Actions de Recherche Concertee de la Communaute Francaise, Belgium. We would like to thank Pierre Gaspard and Thierry Visart de Bocarme for useful discussions. Computational resources have been provided by the Consortium des Équipements de Calcul Intensif (CÉCI), funded by the Fonds de la Recherche Scientifique de Belgique (F.R.S.-FNRS) under Grant No. 2.5020.11.

% Produces the bibliography via BibTeX.

\end{document}